\documentclass[journal, onecolumn]{IEEEtran}

\usepackage{graphicx}    
\usepackage{float}
\usepackage{epstopdf} 
\usepackage{cite}  
\usepackage[labelfont=bf]{caption}
\usepackage{subcaption}

\usepackage{amsmath} 
\usepackage{amssymb} 
\usepackage{amsfonts}

\usepackage{color}

\newtheorem{assumption}{Assumption}
\newtheorem{remark}{Remark}
\newtheorem{lem}{Lemma}
\newtheorem{prop}{Proposition}
\newtheorem{pf}{Proof}

\newcommand{\overbar}[1]{\mkern 1.5mu\overline{\mkern-1.5mu#1\mkern-1.5mu}\mkern 1.5mu}

\newcommand{\rea}{\mathrm{R}}
\newcommand{\img}{\mathrm{i}}
\newcommand{\col}[1]{\mathrm{col}(#1)}
\newcommand{\iinb}{i\in\overbar{N}}
\newcommand{\hth}{\hat{\theta}}
\newcommand{\tth}{\tilde{\theta}}
    
\begin{document}

\title{Improved Transients in Multiple Frequencies Estimation via Dynamic Regressor Extension and Mixing} 
\author{Stanislav~Aranovskiy$^{1,2}$,
        Alexey~Bobtsov$^{2}$,
				Romeo~Ortega$^{3}$,
        Anton~Pyrkin$^{2}$
\thanks{$^{1}$Stanislav Aranovskiy is with the NON-A team, INRIA-LNE, Parc Scientifique de la Haute Borne 40, avenue Halley Bat.A, Park Plaza, 59650 Villeneuve d'Ascq, France}		
\thanks{$^{2}$Stanislav Aranovskiy, Alexey Bobtsov and Anton Pyrkin are with the Department of Control Systems and Informatics, ITMO University, Kronverkskiy av. 49, Saint Petersburg, 197101, Russia :  {\tt\small aranovskiysv@niuitmo.ru}}
\thanks{$^{3}$Romeo Ortega is with the LSS-Supelec, 3, Rue Joliot-Curie, 91192 Gif--sur--Yvette, France : {\tt\small ortega@lss.supelec.fr}}
}

\maketitle 

\begin{abstract}                
A problem of performance enhancement for multiple frequencies estimation is studied. First, we consider a basic gradient-based estimation approach with global exponential convergence. Next, we apply dynamic regressor extension and mixing technique to improve transient performance of the basic approach and ensure non-strict monotonicity of estimation errors. Simulation results illustrate benefits of the proposed solution.
\end{abstract}


\section{Introduction}
A problem of frequency identification for sinusoidal signals attracts researchers' attention both in control and signal processing communities due to its practical importance. Indeed, frequency identification methods are widely used in fault detection systems \cite{goupil2010oscillatory}, for periodic disturbance attenuation \cite{landau2011adaptive, bobtsov2011compensation}, in naval applications \cite{belleter2013globally} and so on.

Many online frequency estimation methods are currently available in literature, e.g. a phase-locked loop (PLL) proposed in \cite{wu2003magnitude}, adaptive notch filters \cite{regalia1991improved, mojiri2004adaptive}. Another popular approach is to find a parametrization yielding a linear regression model, which parameters are further identified with pertinent estimation techniques, see \cite{xia, chen2014robust, fedele2014frequency}. However, the most of online methods are focused on stability studies and local or global convergence analysis; transients performance is not usually considered and is only demonstrated with simulations. On the other hand, it is well-known that many gradient-based estimation methods can exhibit poor transients even for relatively small number of estimated parameters, the transients can oscillate or even display a peaking phenomena. A method to increase frequency estimation performance with adaptive band-pass filters was proposed in \cite{Aranovskiy2015Cascade} but for a single frequency case only. Thus, the problem of performance improvement for multiple frequencies estimation remains open.

A novel way to improve transient performance for linear regression parameters estimation was proposed in \cite{Aranovskiy2015DREM}; the approach is based on extension and mixing of the original vector regression in order to obtain a set of scalar equations. In this paper we apply this approach to the problem of multiple frequencies estimation. It is shown that under some reasonable assumptions and neglecting fast-decaying terms, \emph{non-strict monotonicty} can be provided for estimates of parameters avoiding any oscillatory or peaking behavior. 

The paper is organized as follows. First, in Section \ref{sec:PS} a multiple frequencies estimation problem is stated. A basic method to solve the problem is presented in Section \ref{sec:basic}. Next, in Section \ref{sec:DREM} we consider dynamic regressor extension and mixing (DREM) procedure and apply it to the previously proposed method. Illustrative results are given in Section \ref{sec:sim} and the paper is wrapped up with Conclusion.

\section{Problem Statement} \label{sec:PS}
Consider the measured scalar signal 
\begin{equation} \label{eq:u}
	u(t) = \sum_{i=1}^N{A_i \sin (\omega_i t + \varphi_i)},
\end{equation}
where $t\ge 0$ is time, $A_i>0$, $\varphi_i \in [0, 2\pi)$, and $\omega_i>0$ are unknown amplitudes, phases, and frequencies, respectively, $\iinb:=\{1, 2, \ldots N\}$, $N$ is the number of the frequencies in the signal. 

\begin{assumption} \label{as:as1}
All the frequencies $\omega_i$, $\iinb$, are distinguished, i.e.
\[
	\omega_i \ne \omega_j \  \forall i\ne j, \  i,j\in\overbar{N}.
\]
\end{assumption}

\begin{remark} \label{rem:theta}
The signal \eqref{eq:u} can be seen as an output of a marginally stable linear signal generator 
\[
	\begin{aligned}
		\dot{z}(t) &= \Gamma z(t), \quad z(0)=z_0\in\rea^{2N},\\
		u(t) &= Hz,
	\end{aligned}
\]
where $\Gamma \in \rea^{2N\times 2N}$ and $H\in \rea^{1\times 2N}$. The characteristic polynomial of the matrix $\Gamma$ is given by 
\begin{equation*}
	P_\theta(s):= s^{2N} + \theta_1 s^{2N-2} + \ldots \theta_{N-1}s^2 + \theta_N,
\end{equation*}
where the parameters $\theta_i$ are such that roots of the polynomial $P_\theta(s)$ are $\pm \img \omega_i$, where $\img:=\sqrt{-1}$, $\iinb$. Obviously, given a vector $\theta:=\col{\theta_i}\in\rea^N$, the frequencies can be univocally (up to numerical procedures accuracy) defined, and \emph{vice versa}. Thus, in many multiple frequencies estimation methods the vector $\theta$ is identified instead of separate frequencies values. In our paper we follow this approach and assume that the frequencies are estimated if the vector $\theta$ is obtained. The problem of \emph{direct} frequency identification is considered, for example, in  \cite{Pin2015Direct}.
\end{remark}

\emph{Frequencies Estimation Problem.} The goal is to find mappings $\Psi:\rea \times \rea^{l} \mapsto \rea^l$ and $\Theta: \rea^l\mapsto\rea^N$, such that the following estimator 
\begin{equation} \label{eq:estim}
	\begin{aligned}
		\dot\chi(t) &= \Psi(\chi(t), u(t)), \\
		\hth(t) &= \Theta(\chi(t)).
	\end{aligned}
\end{equation}
ensures
\begin{equation} \label{eq:goal}
	\lim_{t\to\infty}|\hth(t)-\theta|=0.
\end{equation}

\section{A Basic frequencies identification method} \label{sec:basic}
In this section we consider a multiple frequencies estimation method, proposed in \cite{aranovskiy2010identification} and further extended in \cite{bobtsov2012switched, pyrkin2015estimation}. This method is based on State-Variable Filter (SVF) approach, see \cite{young1981parameter, Garnier2003SVF}. 

\begin{lem} \label{lem:svf}
Consider the following SVF
\begin{equation} \label{eq:SVF}
	\dot{\xi}(t) = A \xi(t) + B u(t),
\end{equation}
where $\xi:=\begin{bmatrix}\xi_1(t), & \xi_2(t), & \ldots & \xi_{2N}(t)\end{bmatrix}^\top$,
\[
	A=\begin{bmatrix}
	0 & 1 & 0 & \ldots & 0 \\ 
	0 & 0 & 1 & \ldots & 0 \\ 
	\vdots & \vdots & \vdots & \ddots & \vdots \\ 
	0 & 0 & 0 & \ldots & 1 \\ 
	-a_0 & -a_1 & -a_2 & \ldots & -a_{2N-1}
	\end{bmatrix},
	\quad
	B=\begin{bmatrix} 0 \\ 	0 \\ \vdots \\ 	0 \\ a_0 \end{bmatrix},
\]
$a_i$, $i\in\{0,2,\ldots,2N-1\}$, are coefficients of a Hurwitz polynomial 
\[
	a(s)=s^{2N} + a_{2N-1} s^{2N-1} + \ldots a_1s + a_0.
\]
Define
\begin{equation} \label{eq:y}
	y(t):=-\dot{\xi}_{2N}(t) = \sum_{i=1}^{2N}{a_{i-1}\xi_i(t)} - a_0u(t) .
\end{equation}
Then the following holds:
\begin{equation} \label{eq:reg}
	y(t) = \phi^\top(t)\theta + \varepsilon(t),
\end{equation}
where 
\begin{equation} \label{eq:phi}
	\phi(t):=\begin{bmatrix} \xi_{2N-1}(t), & \xi_{2N-3}(t),  & \ldots & \xi_3(t) & \xi_1(t) \end{bmatrix}^\top,
\end{equation}
$\theta$ is defined in Remark \ref{rem:theta}, and $\varepsilon(t)$ is an exponentially decaying term. 
\end{lem}
The proof is straightforward and follows the proof presented in \cite{pyrkin2015estimation}.

Using Lemma \ref{lem:svf} we can propose a multiple frequencies estimator.

\begin{prop} \label{prop:estim}
Consider the signal \eqref{eq:u} satisfying Assumption \ref{as:as1}, the SVF \eqref{eq:SVF}, and the signals $y(t)$ and $\phi(t)$, defined by \eqref{eq:y} and \eqref{eq:phi}, respectively. Then the estimator
\begin{equation} \label{eq:hattheta}
	\dot{\hth}(t) = K_\theta\phi(t)(y(t) - \phi^\top(t)\hth(t)),
\end{equation}
where $K_\theta\in\rea^{N\times N}$, $K_\theta>0$, ensures the goal \eqref{eq:goal}. Moreover, the estimation error $\tth(t):=\hth(t)-\theta$ converges to zero exponentially fast. 
\end{prop}

\begin{remark}
	The proposed estimator can be also written in form \eqref{eq:estim} as (the argument of time is omitted):
	\begin{align*}
		\chi &:= \col{\xi, \hth}, \\
		\Psi(\chi, u) &:= \begin{bmatrix}
		A\xi+Bu \\ 
		K_\theta\phi(y - \phi^\top\hth)
		\end{bmatrix}, \\
		\Theta(\chi) &:= \hth.
	\end{align*}
\end{remark}

\emph{Sketch of the proof.} The proof of Proposition \ref{prop:estim} follows the proof given in \cite{pyrkin2015estimation}. Substituting \eqref{eq:reg}, it is easy to show that the estimation error $\tth(t)$ obeys the following differential equation
\begin{equation} \label{eq:tildtheta}
	\dot{\tth}(t) = -K_\theta \phi(t)\phi^\top(t)\tth(t) + \epsilon(t),
\end{equation}
where $\epsilon(t):=K_\theta\phi(t)\varepsilon(t)$ is bounded and exponentially decays. Since signal \eqref{eq:u} consists of $N$ sinusoidal components with distinguished frequencies, the vector $\phi(t)$ satisfies \emph{persistant excitation} condition \cite{sastry2011adaptive}, that is 
\[
	\int_t^{t+T} \phi(s) \phi^\top(s)ds \geq \delta I_q,
\]
for some $T,\delta >0$ and for all $t \geq 0$, which will be denoted as $\phi(t) \in \mbox{PE}$. Thus, the linear time-varying system \eqref{eq:tildtheta} is exponentially stable and 
\[
	\lim_{t\to\infty}|\tth(t)|=0. 
\] 

The estimation algorithm \eqref{eq:hattheta} ensures global exponential convergence of $\tth(t)$, but do not guarantee performance transients. It is known from practice that for $N\ge2$ behavior of the estimator \eqref{eq:hattheta} becomes oscillatory and can exhibit peaking phenomena. However, these limitations can be overcome with DREM technique presented in the next section.

\section{Enchancing the basic algorithm via DREM procedure} \label{sec:DREM}
In this section we first present the DREM procedure proposed in \cite{Aranovskiy2015DREM}, and then apply it to the basic frequencies estimation algorithm studied in Section \ref{sec:basic}.

\subsection{Dynamic Regressor Extension and Mixing}
 Consider the basic linear regression 
\begin{equation}\label{eq:regrho}
	\rho(t) = m^\top(t)r,
\end{equation}
where $\rho\in\rea$ and $m \in \rea^q$ are measurable bounded signals and $r\in\rea^q$ is the vector of unknown constant parameters to be estimated. The standard gradient estimator, equivalent to \eqref{eq:hattheta},
\begin{equation*}
	\dot {\hat r}(t)=K_r m(t) (\rho(t) - m^\top(t) \hat{r}(t)),
\end{equation*}
with a positive definite adaptation gain $K_r \in \rea^{q \times q}$ yields the error equation
\begin{equation}  \label{eq:tilder}
	\dot {\tilde r}(t)=-K_r m(t) m^\top(t) \tilde r(t),
\end{equation}
where $\tilde{r}(t) := \hat{r}(t)-r$ is the parameters estimation error.

We propose the following dynamic regressor extension and mixing procedure. The first step in DREM is to introduce $q-1$ {\em linear, $\mathcal{L}_\infty$-stable} operators $H_i: \mathcal{L}_\infty \to \mathcal{L}_\infty,\;i \in \{1,2,\dots,q-1\}$, whose output, for any bounded input, may be decomposed as
\begin{equation*} \label{eq:H}
	(\cdot)_{f_i}(t):=[H_i(\cdot)](t) + \epsilon_t,
\end{equation*}
with $\epsilon_t$ is a (generic) exponentially decaying term. For instance, the operators $H_i$ may be simple, exponentially stable {\em LTI filters} of the form
\begin{equation*}
	H_i(p)=\frac{\alpha_i}{p + \beta_i},
\end{equation*}
with $\alpha_i\ne 0$, $\beta_i>0$; in this case $\epsilon_t$ accounts for the effect of initial conditions of the filters. Another option of interest are {\em delay operators}, that is
\[
	[H_i(\cdot)](t):=(\cdot)(t-d_i),
\]
where $d_i>0$.

Now, we apply these operators to the regressor equation \eqref{eq:regrho} to get the filtered regression\footnote{To simplify the presentation in the sequel we will neglect the $\epsilon_t$ terms. However, it is incorporated in the analysis and proofs given in \cite{Aranovskiy2015DREM}.}
\begin{equation*}
	\rho_{f_i}(t)= m^\top_{f_i}(t) r.
\end{equation*}

Combining the original regressor equation \eqref{eq:regrho} with the $q-1$ filtered regressors we can construct the extended regressor system
\begin{equation} \label{eq:RM}
	R_e(t) = M_e(t) r,
\end{equation}
where $R_e \in \rea^q$ and $M_e \in \rea^{q \times q}$ are defined as
\begin{equation} \label{eq:ReMe}
	R_e(t):=	\begin{bmatrix} \rho(t) \\ \rho_{f_1}(t) \\ \vdots \\ \rho_{f_{q-1}}(t) \end{bmatrix},
\;
M_e(t):=\begin{bmatrix} m^\top(t) \\ m^\top_{f_1}(t) \\ \vdots \\ m^\top_{f_{q-1}}(t) \end{bmatrix}.
\end{equation}
Note that, because of the $\mathcal{L}_\infty$--stability assumption of $H_i$, $R_e$ and $M_e$ are bounded. Premultiplying \eqref{eq:RM} by the {\em adjunct matrix} of $M_e$ we get $q$ scalar regressions of the form 
\begin{equation} \label{eq:Rscalar}
	R_i(t) = \psi_m(t) r_i
\end{equation}
with $i \in \bar q:= \{1,2,\dots,q\}$, where we defined the determinant of $M_e$ as
\begin{equation} \label{eq:psim}
\psi_m(t):=\det \{M_e(t)\},
\end{equation}
and the vector $R \in \rea^q$
\begin{equation} \label{eq:R}
	R(t) := \mathrm{adj}\{M_e(t)\} R_e(t).
\end{equation}

\begin{prop} \label{prop:DREM}
Consider the $q$--dimensional linear regression \eqref{eq:regrho}, where $\rho(t)$ and $m(t)$ are known, bounded functions of time and $r \in \rea^q$ is a vector of unknown parameters. Introduce  $q-1$ linear, $\mathcal{L}_\infty$--stable operators $H_i: \mathcal{L}_\infty \to \mathcal{L}_\infty,\;i \in \{1,2,\dots,q-1\}$ verifying \eqref{eq:H}. Define the vector $R_e$ and the matrix $M_e$ as given in \eqref{eq:ReMe}. 
Next consider the estimator
\begin{equation}
	\dot{\hat{r}}_i = k_i \psi_m(t) (R_i(t) - \psi_m(t) \hat{r}_i) ,\;i \in \bar q,
\end{equation}
where $k_i>0$, $\psi_m(t)$ and $R(t)$ are defined in \eqref{eq:psim} and \eqref{eq:R}, respectively. The following implications holds:
\begin{equation} \label{eq:notL2}
	\psi_m(t) \notin \mathcal{L}_2\quad \Longrightarrow \quad \lim_{t\to \infty} \tilde r_i(t)=0, \ \forall i\in\bar{q}.
\end{equation}
Moreover, if $\psi_m(t) \in \mbox{PE}$, then $\tilde r_i(t)$ tends to zero exponentially fast. 
\end{prop}

\begin{remark}
It is well--known \cite{sastry2011adaptive} that the zero equilibrium of the linear time--varying system \eqref{eq:tilder} is (uniformly) exponentially stable if and only if the regressor vector $m(t)\in\mbox{PE}$. However, the implication \eqref{eq:notL2} proposes a novel criterion for \emph{assymptotic} convergence which is not necessary uniform for $\psi_m(t) \notin \mbox{PE}$. This criterion, namely $	\psi_m(t) \notin \mathcal{L}_2$, is established not for the regressor $m(t)$ itself, but for a determinant of the extended matrix $M_e$, and do not coincide with the condition $m(t)\in\mbox{PE}$. For more details and illustrative examples see \cite{Aranovskiy2015DREM}.
\end{remark}

\begin{remark} \label{rem:mono}
It is easy to show that error dynamics is given by
\[
	\dot{\tilde {r}}_i(t) = -k_i \psi_m^2(t)\tilde{r}_i(t).
\]
It follows that all the transients are \emph{non-strictly monotonic} and do not exhibit oscillatory behavior. 
\end{remark}

\subsection{Applying DREM for frequencies estimation}

Following the proposed procedure we introduce $N-1$ linear, $\mathcal{L}_\infty$--stable delay operators $[H_i(\cdot)](t):=(\cdot)(t-d_i)$,  $i\in \{1, 2, \ldots N-1\}$, where $d_i>0$ and $d_i \ne d_j$ for $i\ne j$, and define $N-1$ filtered signals
\begin{equation} \label{eq:filtphiy}
	\begin{aligned}
		\phi_{f_i}(t) &= \phi(t-d_i), \\
		y_{f_i}(t) &= y(t-d_i).
	\end{aligned}
\end{equation}
Next coupling these signals with $y(t)$ and $\phi(t)$ we construct
\begin{equation} \label{eq:YePhie}
	Y_e(t):=	\begin{bmatrix} y(t) \\ y_{f_1}(t) \\ \vdots \\ y_{f_{N-1}}(t) \end{bmatrix},
\;
	\Phi_e(t):=\begin{bmatrix} \phi^\top(t) \\ \phi^\top_{f_1}(t) \\ \vdots \\ \phi^\top_{f_{N-1}}(t) \end{bmatrix},
\end{equation}	
where $Y_e(t)$ is a $N\times 1$ vector and $\Phi_e(t)$ is a $N \times N$ matrix. Defining 
\begin{equation} \label{eq:psiphi}
	\psi_\phi(t) := \det\{\Phi_e(t)\}
\end{equation}
and
\begin{equation} \label{eq:Yphi}
	Y(t)=\mathrm{adj}\{\Phi_e(t)\} Y_e(t),
\end{equation}
we result with a set of $N$ scalar equations
\[
	Y_i(t) = \psi_\phi(t) \theta_i.
\]
Next the basic differentiator $\eqref{eq:hattheta}$ is replaced with
\begin{equation} \label{eq:estimdrem}
	\dot{\hth}_i(t) = \gamma_i \psi_\phi(t)(Y_i(t)-\psi_\phi(t)\hth(t)),
\end{equation}
where $\gamma_i>0$, $\iinb$.

Following the Proposition \ref{prop:DREM} and Remark \ref{rem:mono}, we are now in position to establish the main result.

\begin{prop} \label{prop:DREMEstim}
Consider the signal \eqref{eq:u} and the SVF \eqref{eq:SVF}. Define $y(t)$ and $\phi(t)$ as \eqref{eq:y} and \eqref{eq:phi}, respectively. Choose $N-1$ parameters $d_i$, $i=\{1, 2, \ldots, N-1\}$ and compute $Y_e(t)$ and $\Phi_e(t)$ as \eqref{eq:filtphiy},\eqref{eq:YePhie}. If the parameters $d_i$ are chosen such that $\psi_\phi(t) \notin \mathcal{L}_2$, where $\psi_\phi(t)$ is defined in \eqref{eq:psiphi}, then  estimation algorithm \eqref{eq:estimdrem} with $Y(t)$ defined in \eqref{eq:Yphi} guarantees for $\iinb$
\begin{itemize}
	\item $\lim_{t\to\infty}{|\hat\theta_i(t)-\theta_i|}=0$;
	\item $\hat{\theta}_i(t)$ is non-strictly monotonic and $|\tth_i(t)|$ is non-increasing.
\end{itemize}
Moreover, if $\psi_\phi(t) \in \mbox{PE}$, then $\hat\theta_i(t)$ converges to $\theta_i$ exponentially fast. 
\end{prop}

The main novelty of Proposition \ref{prop:DREMEstim} in compare with the basic algorithm given in Proposition \ref{prop:estim} consists in guaranteed non-strict monotonicity of the transients $\hat{\theta}_i(t)$. Obviously, the second statement of Proposition \ref{prop:DREMEstim} is only valid neglecting exponentially decaying terms in SVF transients, namely $\varepsilon(t)$ in \eqref{eq:reg}. However, these transients depend on our choice of SVF matrix $A$ in \eqref{eq:SVF}, and, practically, are significantly faster then the estimation process.
\section{An Example} \label{sec:sim}
As an illustrative example we consider the case $N=2$, i.e.
\begin{equation} \label{eq:uN2}
	u(t) = A_1\sin(\omega_1t+\varphi_1) + A_2\sin(\omega_2 t +\varphi_2).
\end{equation}
First we are to choose the tuning parameters
\begin{itemize}
\item SVF \eqref{eq:SVF} with the characteristic polynomial of the matrix $A$
\[
	a(s) = \left(s+\lambda\right)^4,
\]
where $\lambda>0$;
\item the linear delay operator $[H_1(\cdot)](t):=(\cdot)(t-d_1)$, where $d_1>0$;
\item the tunning gains $\gamma_{1,2}>0$. 
\end{itemize}

Next we construct $\phi(t) =[\xi_3(t),\; \xi_1(t)]^\top$,  ${\phi_{f_1}(t) =\phi(t-d_1)}$, $y(t)$ as \eqref{eq:y}, $y_{f_1}(t) =y(t-d_1)$, and the matrices
\begin{equation}\label{eq:PhieN2}
	\Phi_e(t) = \begin{bmatrix}
	\xi_3(t) & \xi_1(t) \\ 
	\xi_3(t-d_1) & \xi_1(t-d_1)
	\end{bmatrix}, 
\end{equation}
\begin{equation*}
Y(t)=\mathrm{adj}\{\Phi_e(t)\} \begin{bmatrix} y(t) \\ y_{f_1}(t)\end{bmatrix}.
\end{equation*}

Applicability of the DREM procedure is stated in the following proposition.
\begin{prop} \label{prop:delN2}
The condition
\begin{equation} \label{eq:N2cond}
	d_1 < \frac{\pi}{\max\{\omega_1,\omega_2\}}
\end{equation}
is sufficient to ensure $\det\{\Phi_e(t)\} \notin \mathcal{L}_2$.
\end{prop}
\begin{remark}
The condition \eqref{eq:N2cond} is not, actually, restrictive, since in many practical scenarios it is reasonable to assume a known upper bound, i.e. $\overbar{\omega}\ge\omega_i$ $\forall \iinb$. Then \eqref{eq:N2cond} is satisfied for $d_1<\pi\overbar{\omega}^{-1}$.
\end{remark}
\begin{pf}
Neglecting exponentially decaying terms, for the states of SVF we have
\[
	\begin{aligned}
		\xi_1(t) &= B_1\sin(\omega_1t+\bar{\varphi}_1) + B_2\sin(\omega_2 t +\bar{\varphi}_2), \\
		\xi_3(t) &= -\left(\omega_1^2 B_1\sin(\omega_1t+\bar{\varphi}_1) + \omega_2^2 B_2\sin(\omega_2 t +\bar{\varphi}_2)\right),
	\end{aligned}
\]
where the parameters $B_{1,2}>0$ and $\bar{\varphi}_{1,2}\in [0, 2\pi)$ depend on the choice of $\lambda$ and parameters of the signal \eqref{eq:uN2}.

Define the function
\[
	I(t) := \int_0^t{\left(\det\{\Phi_e(s)\}\right)^2 ds}.
\]
Tedious but straightforward trigonometric computations yield
\[
		I(t) = C_{lin}t + C_{per}(t) + C_0,
\] 
where the \emph{linear} term is
\[
	C_{lin}:=\frac{1}{2}B_1^2 B_2^2 (\omega_1^2-\omega_2^2)^2 \left(1-\cos(d_1\omega_1)\cos(d_1\omega_2)\right),
\]
$C_{per}(t)$ is a bounded periodic term and $C_0$ is a constant. The condition $\det\{\Phi_e(t)\} \notin \mathcal{L}_2$ is equivalent to $I(t) \to \infty$ as $t \to \infty$, that is satisfied if $C_{lin}\ne 0$. Noting that 
\[
	|\cos(d_1\omega_i)|<1, \ i=1,2
\]
follows from \eqref{eq:N2cond} and recalling Assumption \ref{as:as1} complete the proof. 
\end{pf}

It is worth noting that condition $C_{lin}\ne 0$ implies that $d_1$ is not a period of the signals $u(t)$, $\xi_i(t)$, or a half of period if the signals have half-wave symmetry, ${u(t-d_1)\ne \pm u(t)}$; otherwise the matrix $\Phi_e(t)$ is singular for all $t\ge0$. The inequality \eqref{eq:N2cond} guarantees that $d_1$ is smaller then a have of the smallest period among sinusoidal components with the frequencies $\omega_{1,2}$; it is sufficient but conservative estimate. 

\begin{figure}[t]
	\centering
	\includegraphics[width=0.48\linewidth]{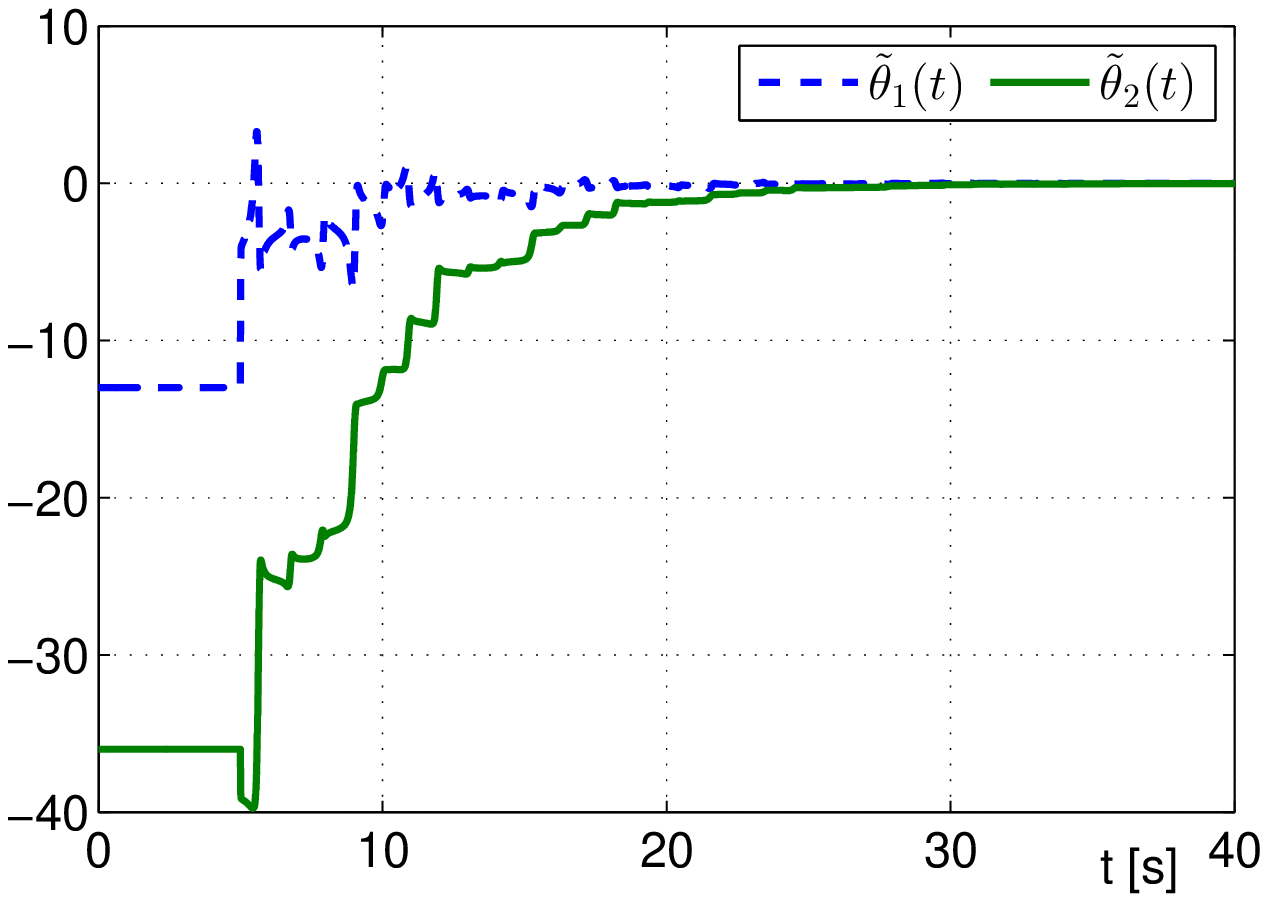}
	\caption{Transients of the basic estimator \eqref{eq:hattheta} for the input signal \eqref{eq:SimU} with $\lambda=5$, $K=\protect\begin{bmatrix} 30 & 0\\0&3\protect\end{bmatrix}$.}
	\label{fig:basic}
	\includegraphics[width=0.48\linewidth]{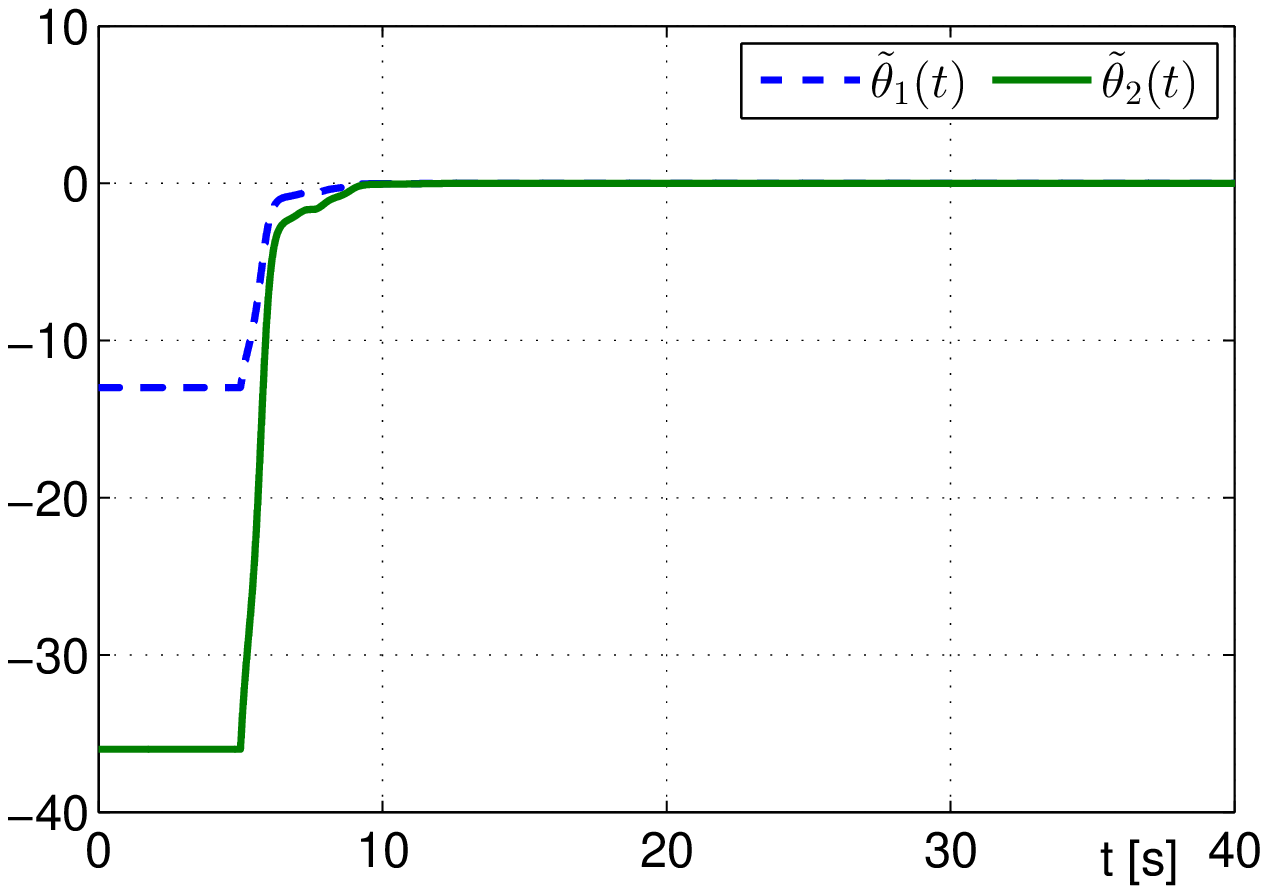}
	\caption{Transients of the estimator with DREM \eqref{eq:estimdrem} for the input signal \eqref{eq:SimU} with $\lambda=5$, $d_1=0.3$, $\gamma_1=\gamma_2=0.1$.}
	\label{fig:drem}
\end{figure}

The both estimators \eqref{eq:hattheta} and \eqref{eq:estimdrem} were simulated for the input signal
\begin{equation}\label{eq:SimU}
	u(t) = 1.2 \sin(2t+\frac{\pi}{3}) + 2\sin(3t+\frac{\pi}{4}).
\end{equation}
with the following parameters
\begin{itemize}
\item $\lambda=5$, $K=\begin{bmatrix} 30 & 0 \\ 0 & 3 \end{bmatrix}$ for the estimator \eqref{eq:hattheta};
\item $\lambda=5$, $d_1=0.3$, $\gamma_1=\gamma_2=0.1$ for the estimator \eqref{eq:estimdrem}.
\end{itemize}
Zero initial conditions are chosen for the both estimators, $\hth(0)=0$, that implies $\tth(0)=-\theta=-[13, \; 36]^\top$. To separate transients of the estimator and of the SVF both the estimators are turned on at $t=5$ seconds.

\begin{figure}[t]
	\centering
\subcaptionbox{$\tth_1(t)$}{\includegraphics[width=0.48\linewidth]{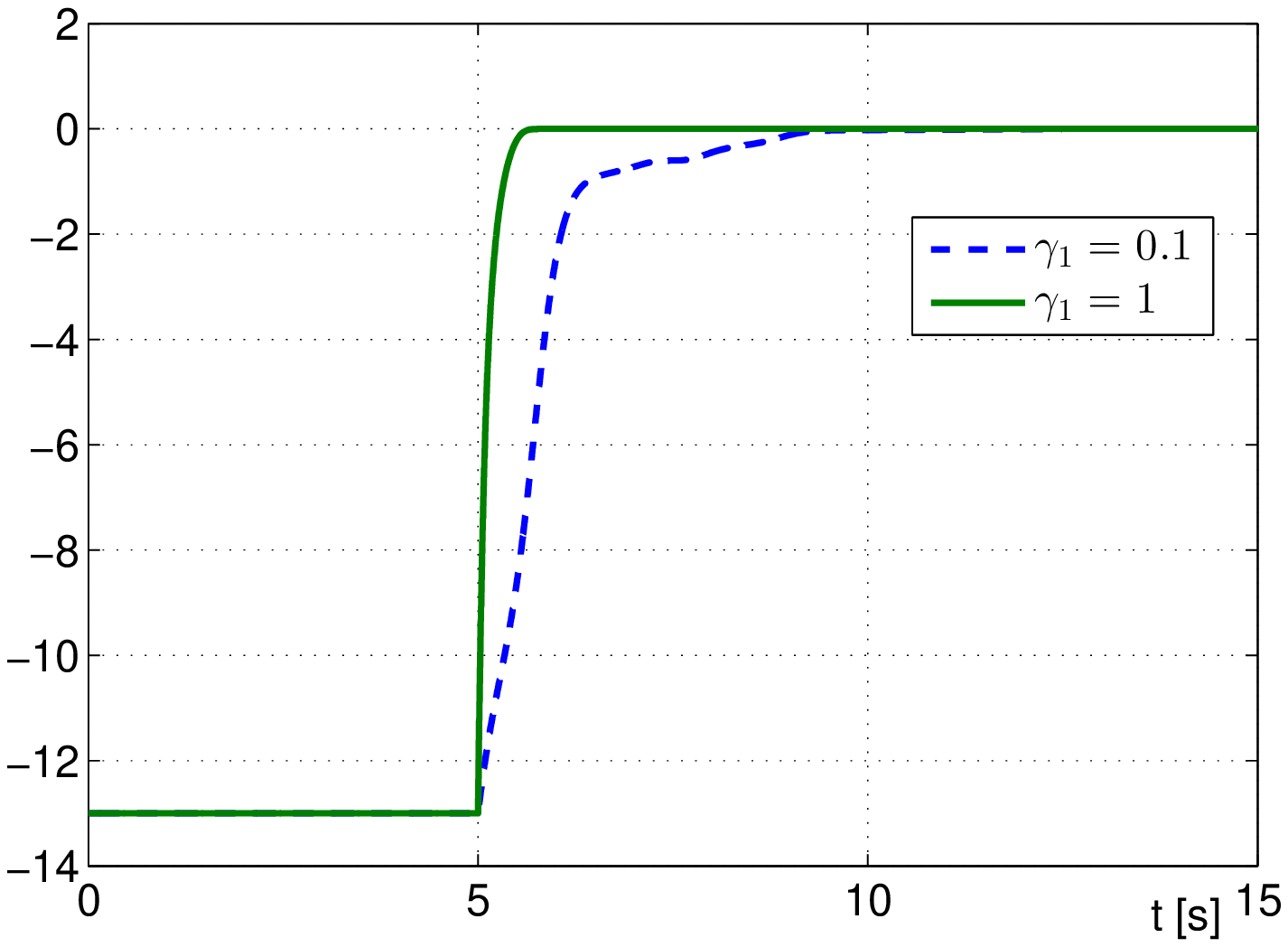}}
\subcaptionbox{$\tth_2(t)$}{\includegraphics[width=0.48\linewidth]{{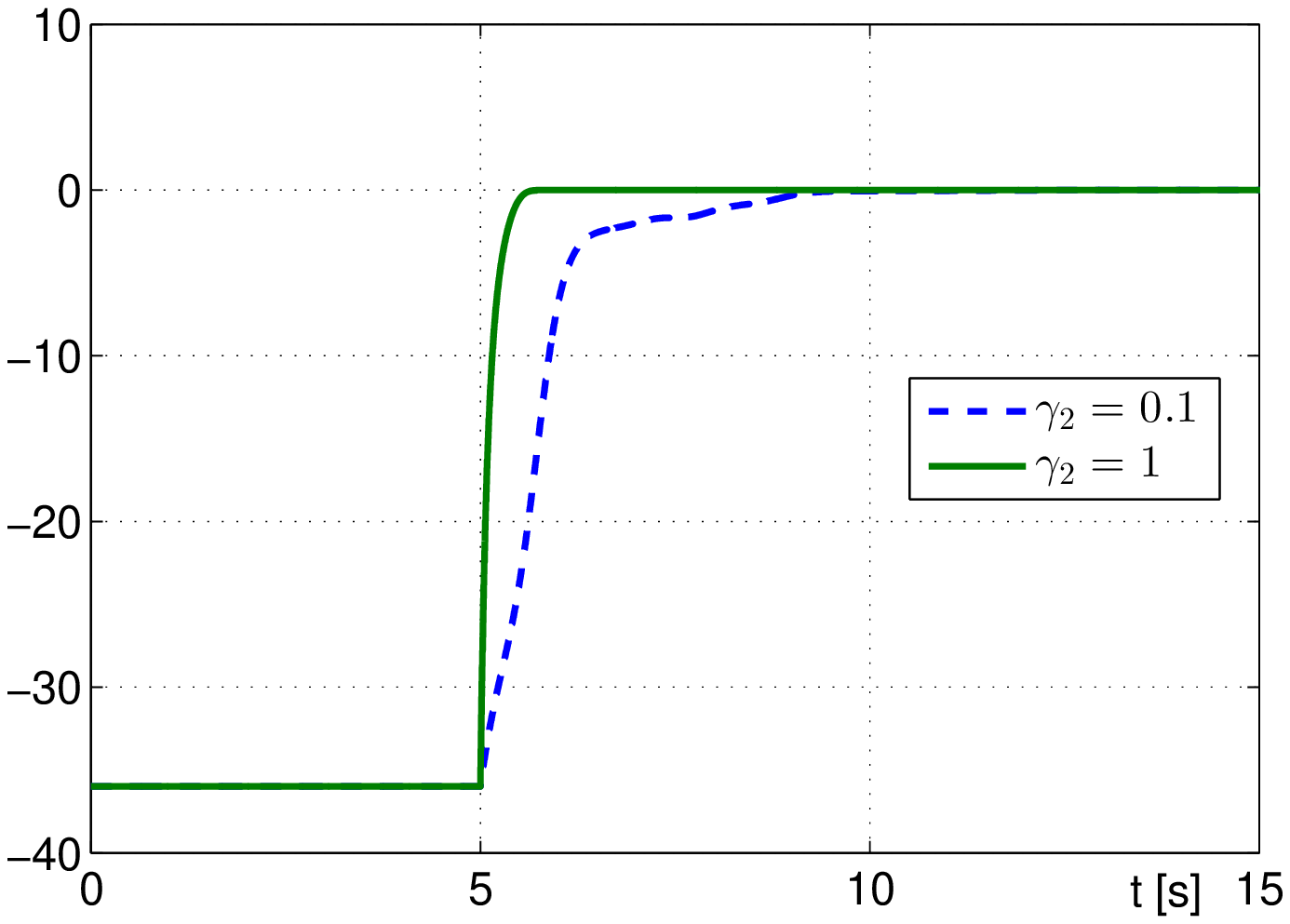}}}
	\caption{Transients comparision of the estimator with DREM \eqref{eq:estimdrem} for the input signal \eqref{eq:uN2} with $\lambda=5$, $d_1=0.3$ and different gains.}
	\label{fig:difgamma}
\end{figure}

Transients $\tth(t)$ of the estimator \eqref{eq:hattheta} are presented in Fig.\ref{fig:basic}, while transients $\tth(t)$ of the estimator \eqref{eq:estimdrem} are presented in Fig.\ref{fig:drem}; note the difference in gains $K$ and $\gamma_{1,2}$ and in transient time. Transients of the estimator \eqref{eq:estimdrem} with $\lambda=5$, $d_1=0.3$ and different values $\gamma_{1,2}$ are given in Fig. \ref{fig:difgamma} and illustrate the impact of the gains.

\begin{figure}[t]
	\centering
	\includegraphics[width=0.48\linewidth]{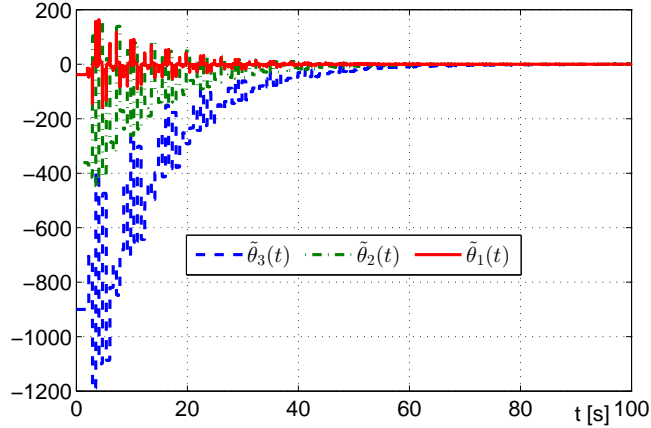}
	\caption{Transients of the basic estimator \eqref{eq:hattheta} for $N=3$ with $\lambda=25$, $K=\protect\begin{bmatrix} 240 &0 &0 \\0 & 40 & 0 \\ 0 & 0 & 10\protect\end{bmatrix}$.}
	\label{fig:N3_1}
\end{figure}

\begin{figure}[t]
	\centering
	\includegraphics[width=0.48\linewidth]{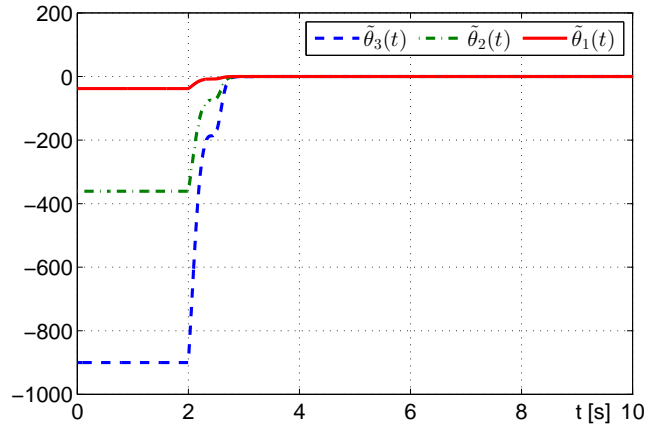}
	\caption{Transients of the estimator with DREM \eqref{eq:estimdrem} for $N=3$ with $\lambda=25$, $d_1=0.2$, $d_2=0.5$, $\gamma_{1}=\gamma_2=\gamma_3=10^{-5}$.}
	\label{fig:N3_2}
\end{figure}

We also present simulation results for $N=3$ and $\theta=[38,\;361,\;900]^\top$ with zero initial condition, SVF \eqref{eq:SVF} with $a(s)=\left(s+\lambda\right)^6$ and $l=25$, and start time $t=2$ seconds. The transients $\tth_{1,2,3}(t)$ are given in Fig. \ref{fig:N3_1} for the estimator \eqref{eq:hattheta} with 
\[
	K=\begin{bmatrix} 240 &0 &0 \\0 & 40 & 0 \\ 0 & 0 & 10\end{bmatrix},
\]
and in Fig. \ref{fig:N3_2} for the estimator \eqref{eq:estimdrem} with $d_1=0.2$, $d_2=0.5$, $\gamma_{1}=\gamma_2=\gamma_3=10^{-5}$; note the difference in time scales.

\section{Conclusion}
The problem of transients improving for multiple frequencies estimation was considered. The dynamic regressor extension and mixing (DREM) procedure, which allows to translate the original vector estimation problem to a set of scalar sub-problems, was successfully applied to enhance the basic estimation algorithm;  as a benefit of this translation the non-strict monotonicity can be ensured. Significant transients improvement is illustrated with simulation results. 

\bibliographystyle{IEEETran}
\bibliography{sin_drem}             

\end{document}